\documentclass[twocolumn,showpacs,showkeys,preprintnumbers,superscriptaddress,amsmath,floatfix,amssymb,secnumarabic]{revtex4}
\usepackage{graphicx}

\newcommand{\comment}[1]{}

\newcommand{\lr}[1]{ \left( #1 \right) }

\newcommand{\lrc}[1]{ \left\{ #1 \right\} }
\newcommand{\vev}[1]{ \langle \, #1 \, \rangle }

\newcommand{\tr}{ {\rm Tr} \, }

\renewcommand{\det}[1]{ {\rm det} \left( #1 \right) }

\newcommand{\sign}{ {\rm sign} \,  }

\newcommand{\expa}[1]{ \exp{\left( #1 \right)} }

\newcommand{\logo}{\\ \vskip -18mm
\leftline{\includegraphics[scale=0.3,clip=false]{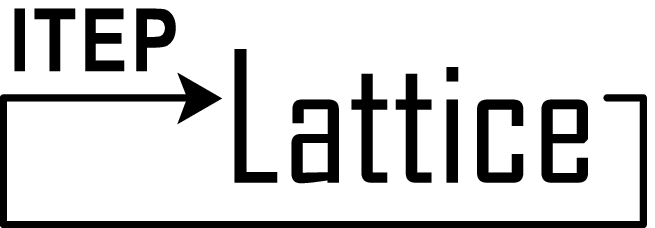}} \vskip 10mm}

\begin{document}
\sloppy
\preprint{ITEP-LAT/2010-11}

\title{A solution of the Gross-Witten matrix model by nonlinear random processes\footnote{Presented at the ``Quark confinement and the hadron spectrum 9'' conference, 29 August - 4 September, Madrid, Spain}\logo}

\author{P. V. Buividovich}
\email{buividovich@itep.ru}
\affiliation{ITEP, 117218 Russia, Moscow, B. Cheremushkinskaya str. 25}
\affiliation{JINR, 141980 Russia, Moscow region, Dubna, Joliot-Curie str. 6}

\date{November 11, 2010}

\begin{abstract}
 We illustrate the stochastic method for solving the Schwinger-Dyson equations in large-$N$ quantum field theories described in \href{http://arxiv.org/abs/1009.4033}{\emph{ArXiv:1009.4033}} on the example of the Gross-Witten unitary matrix model. In the strong-coupling limit, this method can be applied directly, while in the weak-coupling limit we change the variables from compact to noncompact ones in order to cast the Schwinger-Dyson equations in the stochastic form. This leads to a new action with an infinite number of higher-order interaction terms. Nevertheless, such an action can be efficiently handled. This suggests the way to apply the method of \href{http://arxiv.org/abs/1009.4033}{\emph{ArXiv:1009.4033}} to field theories with $U\lr{N}$ field variables as well as to effective field theories in the large-$N$ limit.
\end{abstract}
\pacs{02.70.-c; 02.50.Ey; 11.15.Pg}
\keywords{simulation methods, planar approximation, lattice field theory, random processes}

\maketitle

\section*{Introduction}

 In many physically interesting situations direct Monte-Carlo sampling of the path integral of the theory is not an efficient simulation method. Probably the most notable example in hadronic physics is QCD at finite chemical potential. Another example are quantum field theories in the large-$N$ limit, since the computational complexity of Monte-Carlo algorithms tends to infinity in this limit, and one always has to rely on extrapolations from finite $N$.

 One of the ways to overcome these difficulties is to expand the desired path integral in powers of some coupling constants, and to sum up the obtained series stochastically. This is the basic idea of the Diagrammatic Monte-Carlo \cite{Prokofev:98:1, Wolff:09:1}. This method, however, cannot be straightforwardly applied to field theories with continuous variables, since in such cases the perturbative series are typically only asymptotic \cite{Prokofev:10:1}.

 For field theories with a non-Abelian internal symmetry group, such as $U\lr{N}$, each diagram contains also some power of $N$, and the sums over the terms leading in $N$ are typically convergent \cite{Brezin:78:1}. This salient feature of large-$N$ quantum field theories was used in \cite{Buividovich:10:2} to construct a stochastic algorithm for summing over all planar diagrams in perturbative expansions. This algorithm is based on the stochastic solution of the Schwinger-Dyson (SD) equations in the factorized form, so that they are interpreted as the equations for the stationary probability distribution of a certain random process. The correlators of the field variables are thus estimated as probability distributions of some random variables, similarly to the ``worm algorithm'' \cite{Prokofev:01:1}. Once the SD equations can be cast in the stochastic form
\begin{eqnarray}
\label{random_process_eq}
w\lr{x} = p_{c}\lr{x} + \sum \limits_{y} p_{e}\lr{x | y} w\lr{y}
 + \nonumber \\
\sum \limits_{y_1, y_2} p_{j}\lr{x | y_1, y_2} w\lr{y_1} w\lr{y_2}  ,
\end{eqnarray}
where $w\lr{x}$ schematically denote the field correlators and the coefficients $p_{c}\lr{x}$, $p_{e}\lr{x | y}$ and $p_{j}\lr{x | y_1, y_2}$ satisfy the inequality $\sum \limits_{x} |p_{c}\lr{x}| + |p_{e}\lr{x|y_1}| + |p_{j}\lr{x | y_1, y_2}| < 1$, one can construct a ``nonlinear'' random process for which $w\lr{x}$ is the stationary probability distribution \cite{Buividovich:10:2}.

 However, when the field variables are the elements of $U\lr{N}$, SD equations can be rewritten as (\ref{random_process_eq}) only in the strong coupling limit. On the other hand, the continuum limit of such theories typically corresponds to the weak-coupling limit. In this paper we consider the simplest nontrivial model defined by the integral over $U\lr{N}$ group in the limit $N \rightarrow \infty$, namely, the Gross-Witten unitary matrix model \cite{Gross:80:1}. We demonstrate that the method of \cite{Buividovich:10:2} can be used to simulate this model in both the strong- and the weak-coupling regimes. In the latter case, one has to change the variables to the noncompact ones, which are more adequate for the weak-coupling expansion.

 The partition function of the Gross-Witten model \cite{Gross:80:1} is
\begin{eqnarray}
\label{gw_pf}
 \mathcal{Z}\lr{\lambda} = \int \limits_{U\lr{N}} dg \, \expa{\frac{N}{\lambda}\, \tr\lr{g + g^{-1}} }  ,
\end{eqnarray}
and the observables of interest are $G_n\lr{\lambda} = \vev{\frac{1}{N}\, \tr g^n}$.

\section*{Strong-coupling regime}

 The SD equations for the model (\ref{gw_pf}) in terms of the observables $G_n\lr{\lambda}$ are:
\begin{eqnarray}
\label{gw_SD_eq}
 G_1 + \frac{1}{\lambda}\, G_{2} - \frac{1}{\lambda} = 0
 \\
\label{gw_sds}
 G_n + \sum \limits_{k=1}^{n-1} G_k \, G_{n-k} + \frac{1}{\lambda}\, G_{n+1} - \frac{1}{\lambda}\, G_{n-1} = 0, \quad n \ge 2
\end{eqnarray}
By a variable redefinition $G_n = \mathcal{N} c^n w_n$ one can transform these equations into the stochastic form (\ref{random_process_eq}) if $\lambda \ge \bar{\lambda}_s \approx 5.94$. According to \cite{Buividovich:10:2}, such equations describe the stationary distribution of the random process operating on the stack of the elements of the form $\lrc{n, \pm}$, where $n \ge 1$ is an integer labeling the observables $G_n$. $w_n$ is proportional to the difference of the probabilities to find the elements $\lrc{n, +}$ and $\lrc{n, -}$ at the top of the stack. At each step of the random process one performs at random one of the following steps:
\begin{description}
  \item[Create]: With probability $\lr{\lambda \mathcal{N} c}^{-1}$ push a new element $\lrc{1, +}$ to the stack.
  \item[Join]: With probability $\mathcal{N}$ pop the two elements $\lrc{m, s_1}$ and $\lrc{n, s_2}$ and push a new element $\lrc{m+n, - s_1 s_2}$.
  \item[Increase]: With probability $\lr{\lambda c}^{-1}$ increase the topmost element in the stack by one.
  \item[Decrease]: With probability $c/\lambda$ decrease the topmost element in the stack by one and change its sign.
  \item[Restart]: Otherwise restart with a stack containing a single element $\lrc{1, +}$.
\end{description}

\begin{figure}
  \includegraphics[width=5cm, angle=-90]{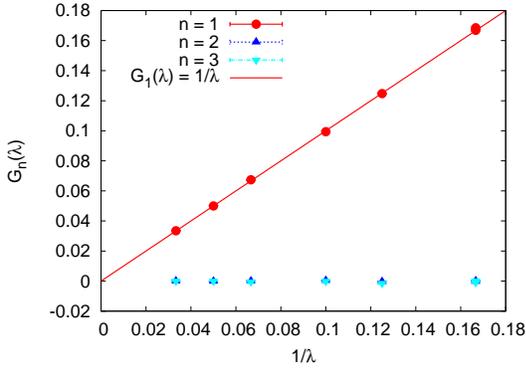}
  \caption{Observables $G_n\lr{\lambda}$ in the strong-coupling regime.}
  \label{fig:gw_sc_seq}
\end{figure}

 The dependence of the observables $G_n\lr{\lambda}$ on $\lambda$ is illustrated on Fig. \ref{fig:gw_sc_seq}. The data were obtained for $10^6$ iterations of the random process, which took several seconds on a standard 2 GHz CPU. In agreement with the exact solution \cite{Gross:80:1}, only $G_1\lr{\lambda}$ is nonzero and behaves as $G_1\lr{\lambda} = \lambda^{-1}$, while all other observables are equal to zero within error range. It should be noted that the element $\lrc{n, \pm}$ with $n > 1$ are generated by our random process, but their contributions to $G_n$ at $n > 1$ cancel on average, and as a result $G_n\lr{\lambda} = 0$ for $n > 1$.

 A similar stochastic strong-coupling solution can be easily constructed for other models with $U\lr{N}$ field variables, for example, for the large-$N$ $U\lr{N}$ non-Abelian lattice gauge theory. In this case the desired random process will be a minor modification of the stochastic solution of the Weingarten model \cite{Buividovich:10:2}. However, such a strong-coupling solution is not very interesting, since the continuum physics should emerge in the weak-coupling limit.

\section*{Weak-coupling regime}

 At sufficiently small $\lambda$ the equations (\ref{gw_SD_eq}) cannot be transformed to the stochastic form by a simple rescaling of observables. In order to obtain a perturbative solution in this case, let us introduce the new Hermitian matrix variables $X$ via the relation $g = \lr{1 + i \alpha X} \lr{1 - i \alpha X}^{-1}$, where the value of $\alpha$ will be given later. First we have to express the Haar measure on $U\lr{N}$ in terms of $X$. It can be found from the invariant metric form $\tr\lr{d g \, d g^{-1}} = g_{AB} dX^A dX^B =  4 \alpha^2 \, \tr\lr{\lr{1 + \alpha^2 X^2}^{-1} d X  \lr{1 + \alpha^2 X^2}^{-1} d X}$:
\begin{eqnarray}
\label{un_inv_metric}
 \sqrt{|\det g_{AB}|} = \expa{-N \tr \log\lr{1 + \alpha^2 X^2}}
\end{eqnarray}
Now one can include the Haar measure (\ref{un_inv_metric}) into the action for the variables $X$ and express integrals over $g$ as integrals over $X$. The new action can be expanded in powers of $X$, thus leading to the Hermitian matrix model with an infinite number of terms in the potential:
\begin{eqnarray}
\label{newvar_action}
S\lr{X}
=
 4 \lambda^{-1} \sum \limits_{k=1}^{+\infty} \lr{-1}^{k-1} \alpha^{2 k} \tr X^{2 k} \, \lr{1 + \frac{\lambda}{4 k}}
\end{eqnarray}
Now we set $\alpha = \lambda/\lr{2 \lambda + 8}$ so that the quadratic term in the action (\ref{newvar_action}) becomes $1/2 \tr X^2$. It is interesting that another parameterizations of $U\lr{N}$ group manifold by Hermitian matrices (e.g. the exponential $g = \expa{i \alpha X}$) can in general lead to multi-trace terms in the action.

\begin{figure}
  \includegraphics[width=5cm, angle=-90]{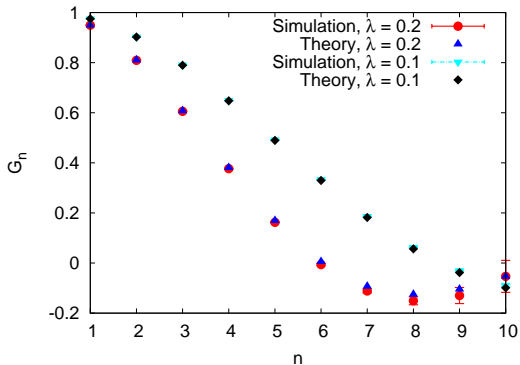}
  \caption{Observables $G_n\lr{\lambda}$ in the weak-coupling phase.}
  \label{fig:gw_wc_seq}
\end{figure}

 The SD equations for the new action (\ref{newvar_action}) involve the observables $\xi_n = \vev{ \frac{1}{N} \, \tr X^{2 n} }$. Again, we rescale the variables as $\xi_n = \mathcal{N} c^n w_n$ to cast the SD equation in the stochastic form (\ref{random_process_eq}):
\begin{eqnarray}
\label{newvar_SD}
w_1 = \lr{\mathcal{N} c}^{-1}
 + \nonumber \\ +
\lambda^{-1} \sum \limits_{k=2}^{+\infty} \lr{-1}^k c^{k-1} \alpha^k \lr{8 k + 2 \lambda} \, w_k
\nonumber \\
w_n = 2/c \, w_{n-1} + \frac{\mathcal{N}}{c}\, \sum \limits_{k=1}^{n-2} w_k \, w_{n-1-k}
+ \nonumber \\ +
\lambda^{-1} \sum \limits_{k=2}^{+\infty} \lr{-1}^k c^{k-1} \alpha^k \lr{8 k + 2 \lambda} \, w_{n + k - 1}
\end{eqnarray}
Following \cite{Buividovich:10:2}, one can devise a random process which samples the unknowns $w_n$. The configuration space of the process is again the stack containing positive integers which label the observables $\xi_n$ and signs. At each time step one performs at random one of the following actions:
\begin{description}
  \item[Create]: With probability $\lr{\mathcal{N} c}^{-1}$ push a new element $\lrc{1, +}$ to the stack.
  \item[Join]: With probability $\mathcal{N}$ pop the two elements $\lrc{m, s_1}$ and $\lrc{n, s_2}$ and push a new element $\lrc{m+n+1, s_1 s_2}$.
  \item[Increase]: With probability $2/c$ increase the topmost element in the stack by one.
  \item[Decrease]: With probability $2 c^k \alpha^{k+1} \lr{4 k + 4 + \lambda}$ decrease the topmost element $\lrc{n, \pm}$ in the stack by $n > k > 0$ and change its sign if $k$ is even.
  \item[Restart]: Otherwise restart with a stack containing a single element $\lrc{1, +}$.
\end{description}
The total probability of all actions can be made less than one if $\lambda \le \bar{\lambda}_w \approx 0.2$.

\begin{figure}
  \includegraphics[width=5cm, angle=-90]{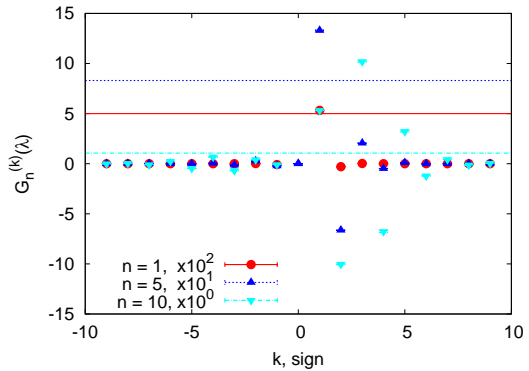}
  \caption{Contributions of different terms of the weak-coupling expansion to $G_n\lr{\bar{\lambda}_w}$, for $10^6$ iterations of the random process.}
  \label{fig:gw_wc_pt}
\end{figure}

 In order to calculate the observables $G_n$, one should expand them in powers of $X$: $\frac{1}{2 N}\, \tr\lr{g\lr{X}^n + g^{-n}\lr{X}} = 1 - \sum \limits_{k = 1}^{+\infty} g_n^{\lr{k}} \frac{1}{N}\, \tr X^{2k}$. $1 - G_n$ can be then found from the averages $\vev{\vev{ \mathcal{N} c^k g_n^{\lr{k}} \sign\lr{k}  }}$ w.r.t. the above described random process. Here $k$ is the topmost element in the stack and $\sign\lr{k}$ is the corresponding sign. The observables $G_n\lr{\lambda}$ are compared with the exact solution \cite{Gross:80:1} on Fig. \ref{fig:gw_wc_seq} for $\lambda = 0.2 \approx \bar{\lambda}_w$ and for $\lambda = 0.1$. In order to illustrate the efficiency of the algorithm, on Fig. \ref{fig:gw_wc_pt} we plot the contributions of different $k$ (positive and negative separately) to $1 - G_n\lr{\lambda}$ at $\lambda = 0.2 \approx \bar{\lambda}_w$. Solid lines are the exact values $G_n\lr{\lambda}$. It should be noted that since $\alpha \sim \lambda$, each such contribution corresponds to the term of order $\lambda^k$ in the weak-coupling expansion of $G_n\lr{\lambda}$. $G_1\lr{\lambda}$ is saturated almost completely by the lowest-order term, while for $G_{10}\lr{\lambda}$ large contributions of different $k$ cancel to yield a numerically small value. Contributions of negative $k$ are numerically small in both cases. Thus the strength of the ``sign problem'' for our random process actually depends on the choice of observables. We also note that no critical slowing down near $\bar{\lambda}_{w,s}$ was observed neither in the weak- nor in the strong-coupling phases.

\begin{acknowledgments}
This work was supported by Grants RFBR 09-02-00338, 10-02-09484-mob\_z, a grant for scientific schools No. NSh-679.2008.2, and by personal grants from the ``Dynasty'' foundation and from the FAIR-Russia Research Center (FRRC).
\end{acknowledgments}

\end{document}